# AFM imaging of SWI/SNF action : mapping the nucleosome remodeling and sliding

Fabien MONTEL, Emeline FONTAINE, Philippe ST-JEAN, Martin CASTELNOVO and Cendrine FAIVRE-MOSKALENKO

Laboratoire Joliot-Curie (CNRS USR 3010) et Laboratoire de Physique (CNRS UMR 5672), Ecole Normale Supérieure de Lyon, 46 Allée d'Italie, 69007 Lyon, France

## ABSTRACT

We propose a combined experimental (Atomic Force Microscopy) and theoretical study of the structural and dynamical properties of nucleosomes. In contrast to biochemical approaches, this method allows to determine simultaneously the DNA complexed length distribution and nucleosome position in various contexts. First, we show that differences in the nucleo-proteic structure observed between conventional H2A and H2A.Bbd variant nucleosomes induce quantitative changes in the in the length distribution of DNA complexed with histones. Then, the sliding action of remodeling complex SWI/SNF is characterized through the evolution of the nucleosome position and wrapped DNA length mapping. Using a linear energetic model for the distribution of DNA complexed length, we extract the net wrapping energy of DNA onto the histone octamer, and compare it to previous studies.

**Keywords :** Atomic Force Microscopy, mono-nucleosome, H2A.Bbd, length distribution of wrapped DNA, nucleosome position distribution, chromatin remodeling factor, histone variant

## INTRODUCTION

DNA is packaged into chromatin in the cell nucleus. The chromatin repeating unit, called the nucleosome, consists of an octamer of the core histones (two each of H2A, H2B, H3 and H4) around which about two superhelical turns of DNA are wrapped (1). The Nucleosome Core Particle (NCP) represents a barrier for the transcription factors binding to their target DNA sequences and interferes with several basic cellular processes (2). Histone modifications, ATP-remodeling machines and the incorporation of histone variants within chromatin are used by the cell to overcome the nucleosomal barrier and modulate DNA accessibility by the control of nucleosome dynamics (3-6). In this work, we use a single molecule technique (Atomic Force Microscopy) to visualize isolated mono-nucleosomes, to quantify the influence of histone octamer composition (H2A-Bbd variant) on the equilibrium nucleosome conformation and to map nucleosome mobility induced by a remodeling complex (SWI/SNF).

Chromatin remodeling complexes are used by the cell to overcome the general repression of transcription associated with the DNA organization into chromatin (7-9). In order to destabilize histone-DNA interaction, remodeling factors (like SWI/SNF) consume the energy from ATP hydrolysis to relocate the histone octamer along the DNA sequence (10, 11) and in some cases, the ejection of the octamer from the DNA template is observed (12). The molecular motor SWI/SNF is known to mobilize the histone octamer from a central to an end-position on short DNA templates (13). Nevertheless, the molecular mechanisms involved in the nucleosome remodeling process have not yet been elucidated .

Histone variants are nonallelic isoforms of the conventional histones. The function of the different histone variants is far from clear, but the emerging general picture suggests that the incorporation of histone variants (14-19) in the nucleosome has serious impacts on several processes, including transcription and repair, and it may have important epigenetic consequences (20-23). H2A.Bbd (Barr body deficient) is an unusual histone variant whose primary sequence shows only 48% identity compared to its conventional H2A counterpart (24). The current view is that H2A.Bbd is enriched in nucleosomes associated with transcriptionally active regions of the genome (24). In recent studies, the unusual properties of this variant nucleosome were described (16, 25) using a combination of physical methods and molecular biology approaches. Those results were mainly focused on the biological role of the various histone fold domains of H2A.Bbd on the overall structure, stability and dynamics of the nucleosome, whereas we concentrate here on the quantification of the subtle modifications in the nucleosome conformation induced by the presence of this histone variant.

Different experimental approaches have been used so far to study the structure and dynamics of the nucleosome, including crystallographic studies by Luger et al. (26) , restriction enzyme accessibility assays (27, 28), and FRET measurements (29, 30). Additionally, physical models (31) and recent computational efforts were developed to describe the nucleosome dynamics and energetics (32-35). Following these numerous contributions, the present study combines experimental (Atomic Force Microscopy) and theoretical tools to bring complementary information regarding the interplay between nucleosome position dynamics and DNA wrapping energetics.

Atomic Force Microscopy (36) allows direct visualization of chromatin fibers and isolated nucleosomes (37). Several experimental procedures allow to depose and observe reproducibly, all this without any fixing agent, DNA or chromatin samples (38-44). By scanning the sample with an apex of very high aspect ratio mounted on a flexible lever, the topography of a surface at the nanometric scale can be acquired. Moreover, computer analysis of AFM images enables the extraction of systematic and statistically relevant distributions of structural parameters describing these biological objects (45-47). As nucleosome is a complex and very dynamic structure, it has been observed that, for a given DNA template, the position of the octamer relative to the sequence (13, 48-50) and the length of DNA wrapped around the histone octamer (27-29, 51, 52) both could change drastically in time.

This paper is organized as follows. First, we show that mapping the nucleosome position along with the length of DNA complexed with histones within individual nucleosome is a powerful tool to discriminate between conventional and variant nucleosomes. A model is then proposed to explain quantitatively these differences and to calculate the wrapping energy of nucleosomes in each case. Next, we have studied nucleosomes in a more dynamic context by observing the action of chromatin remodeling factor SWI/SNF. To do so, similar mapping of the nucleosome position and DNA complexed length was used to quantify the impact of ATP-activated remodeling and sliding of nucleosomes. The results suggest experimental insights into the processivity of SWI/SNF on mono-nucleosomes.

# MATERIALS AND METHODS

**Preparation of DNA fragments**

The 255 bp and 356 bp DNA fragments, containing the 601 nucleosome positioning sequence(53), were obtained by PCR amplification from plasmid pGem-3Z-601. For the 255 bp template, 147 bp long 601 positioning sequence is flanked by 52 bp on one side and 56 bp on the other side. For the 356 bp template, 147 bp long-601 positioning sequence is flanked by 127 bp on one side and 82 bp on the other side. As both 601 DNA templates are built from the same plasmid, the DNA flanking sequences of the short template are included in the long DNA template.

**Protein purification, nucleosome reconstitution and remodeling**

Recombinant Xenopus laevis full-length histone proteins were produced in bacteria and purified as described (54). For the H2A.Bbd protein, the coding sequences for the H2A and for H2A.Bbd were amplified by PCR and introduced in the pET3a vector. Recombinant proteins were purified as previously described (55).

Yeast SWI/SNF complex was purified as described previously (56) and its activity was normalized by measuring its effect on the sliding of conventional nucleosomes : 1 unit being defined as the amount of ySWI/SNF required to mobilize 50% of input nucleosomes (~50 ng) at 29°C during 45 minutes. Nucleosome reconstitution was performed by the salt dialysis procedure (57). Nucleosomes reconstituted on a 601 nucleosome positioning sequence (20 ng) were incubated with SWI/SNF as indicated at 29°C and in remodeling buffer (RB) containing 10 mM Tris-HCl, pH = 7.4, 2.5 mM $MgCl_2$, and 1 mM ATP. The reaction was stopped after the time as indicated by diluting about 10 times in TE buffer (Tris-HCl 10 mM, pH = 7.4, EDTA 1 mM) and NaCl 2 mM and deposing the sample onto the functionalized APTES-mica surface.

**Atomic Force Microscopy and surface preparation**

For the AFM imaging the conventional and variant nucleosomes were immobilized onto APTES-mica surfaces. The functionalization of freshly cleaved mica disks (muscovite mica, grade V-1, SPI) was obtained by self-assembly of a monolayer of APTES under Argon atmosphere for 2 hours (39). Nucleosomes (DNA concentration ~ 75 ng/µl) were filtered and concentrated using Microcon® centrifugal filters to remove free histones from the solution, and diluted 10 times in TE buffer, just prior to deposition onto APTES-Mica surfaces. A 5 µl droplet of the nucleosome solution is applied on the surface for 1 min, rinsed with 1 mL of milliQ-Ultrapure © water and gently dried by nitrogen flow. The samples were visualized by using a Nanoscope III AFM (Digital Instruments™, Veeco, Santa Barbara, CA). The images were obtained in Tapping Mode in air, with silicon tips (resonant frequency 250-350 kHz) or Diamond Like Carbon Spikes tips (resonant frequency ~150 kHz) at scanning rates of 2 Hz over scan areas of 1 µm wide.

This surface functionalization was chosen because it is known to trap 3D conformation of naked DNA molecule on a 2D surface (58, 59). Moreover, under such experimental conditions, rinsing and drying are thought to have little effect on the observed conformation of biomolecule (60).

**Image analysis**

We have extracted parameters of interest from the AFM images using a MATLAB© (The Mathworks, Natick, MA) script essentially based on morphological tools such as binary dilatation and erosion (61-64), and height/areas selections . The aim of the first three steps of this algorithm is to select relevant objects :

1. In order to remove the piezoelectric scanner thermal drift, flatten of the image is performed. The use of a height criteria (h>0.5nm where h is the height of the object) allows to avoid the shadow artifact induced by high objects on the image.
2. Building of a binary image using a simple thresholding (h > 0.25 nm where h is the height of the object)) and then selection of the binary objects in the good area range (500 < A < 2000 nm² where A is the area of the object)).
3. *Selection of the objects in the good height range using a hysteresis thresholding (65) ($h^{min}_1 = 0.25$ nm and $h^{min}_2 = 1.4$ nm, where $h^{min}_1$ and $h^{min}_2$ are the height of the two thresholds).*

These three steps leads to the selection of binary objects whose area is between 500 and 2000 nm² and corresponds in the AFM image to a group of connected pixels whose minimun height is more than 0.25 nm and maximum height is above 1.6 nm. For example a height criterion is used to reject tetrasomes while events with SWI/SNF still complexed with nucleosomes are removed from analysis by a size criterion. The next steps correspond to measurements in itself :

4. Detection of the NCP centroid by shrinking the objects in the binary image.
5. Building of a distance map inside the nucleosome with respect to their NCP centroid using a pseudo-euclidian dilatation based algorithm.
6. Selection of the non-octamer parts of the nucleosomes ($d > d_c$ , where d is the constraint distance to the NCP centroid and $d_c \sim 7.5$ nm is the apparent nucleosome radius) and then thinning of the free arm regions using a commercial MATLAB© script optimised to avoid most of the branching in the skeleton.
7. Selection of the free arm ends and measurement of the free arm lengths.
8. Measurement of other parameters of interest like areas, volumes and mean height of the nucleosomes and the octamers (see *supplemental materials*).

These last 5 steps lead to quick and robust measurements. Indeed the use of morphological tools allows  parallel calculation simultaneously on all the objects. Moreover, erosion is a good approximation for the inverse operation of the AFM dilatation due to the finite tip radius and leads to a partial removal of the tip effect (66, 67).

The longest arm is named $L_+$ and the shortest $L_-$. DNA complexed length is deduced by $L_c = L_{tot} - L_- - L_+$ where $L_{tot}$ is either 255 bp for short conventional and variant nucleosomes or 356 bp for long conventional nucleosomes. The position of the nucleosome relatively to the DNA template center is calculated as $\Delta L = (L_+ - L_-)/2$. Notice that the position defined this way corresponds to the location of the most deeply buried base pair, which might differ from dyad axis position (strictly defined for symmetric nucleosomes).

**Complexed DNA length and nucleosome position distribution construction**

For the distribution of DNA complexed length, well centered nucleosomes were selected ($\Delta L^* - \sigma_{\Delta L}/2 \sim 0$ bp $< \Delta L < 12$ bp $\sim \Delta L^* + \sigma_{\Delta L}/2$ for the 255 bp mono-nucleosomes where $\Delta L^*$ is the most probable nucleosome position and $\sigma_{\Delta L}$ is the standard deviation of the $\Delta L$ distribution). To construct the histogram a 20 bp-sliding box was used. For each $L_0$ in [0, 300 bp], nucleosomes with a DNA complexed length included in the range [$L_0 - 10$ bp, $L_0 + 10$ bp] were counted. After normalization, a smooth distribution is obtained that represents mathematically the convolution of the real experimental distribution with a rectangular pulse of 20 bp long.

To obtain the nucleosome position distribution we have selected nucleosomes with a DNA complexed length $L_c$ in a range of width $\sigma_{Lc}$ around $L^* = 146$ bp (123 bp ~ $L^*$ - $\sigma_{Lc}$ < $L_c$ < $L^*$ - $\sigma_{Lc}$ 169 bp for canonical nucleosomes). Then, the same 20-bp sliding box protocol was used to construct the nucleosome position distribution. The error on the distribution function mean value (standard error) is given by $\sigma_{exp}/\sqrt{N}$, where $\sigma_{exp}$ is the standard deviation of the experimental distribution, and N the number of analyzed nucleosomes (central limit theorem).

## 2D distribution $L_c/\Delta L$ construction

To construct the 2D-histogram a 10 bp-sliding box was used. For each coordinates $(\Delta L_0, L_0)$ in $[0, 75$ bp$]\times[0, 300$ bp$]$, nucleosomes with a DNA complexed length included in the range $[L_0 - 5$ bp, $L_0 + 5$ bp$]$ and a position included in the range $[\Delta L_0 - 5$ bp, $\Delta L_0 + 5$ bp$]$ were counted. After normalization a smooth distribution is obtained that represents mathematically the convolution of the real experimental 2D-distribution with a 10 bp square rectangular pulse.

## Reproducibility and experimental errors

We have checked that different batches of APTES, nucleosome reconstitutions, ySWI/SNF and mica surfaces lead to similar results for the sliding assays and for the 2D mapping within the experimental uncertainty. Moreover we have checked by image analysis of the same naked DNA on the same surface and within the same experimental conditions (data not shown) that the whole measurement and analysis process have an experimental error of about 10 bp in DNA length measurement. Notice that uncertainty on the mean value of length measurements can be much smaller than this resolution as it is explained in the supplemental material S3.

# RESULTS AND DISCUSSION

## Simultaneous measurements of DNA complexed length and nucleosome position.

Several biochemical approaches allow accessing either the nucleosome position along a DNA template, or the length of DNA wrapped around the histone octamer, but using AFM, we were able to measure them simultaneously. The results are conveniently plotted as 2D histograms of nucleosome position versus DNA complexed length.

### *For short and long arm mononucleosomes*

We first investigated the influence of the DNA template length on the nucleosome complexed length distribution for conventional nucleosomes. Indeed, one could expect that the nucleosome positioning efficiency for the 601 DNA template and/or the range of wrapped DNA length could depend on the length of free DNA arms. Using purified conventional recombinant histones, nucleosomes were reconstituted by salt dialysis on 255 bp (short nucleosomes) or 356 bp (long nucleosomes) DNA fragments containing the 601 positioning sequence. Tapping Mode AFM in air was used to visualize the reconstituted particles adsorbed on APTES-mica surfaces and images of 1 µm² were recorded. A representative image of long mono-nucleosomes ($L_{tot}$ = 356 bp) is displayed on Figure 1a. Such an image enables to clearly distinguish the nucleosome core particle (red part of the complex : $h_{NCP} \sim 2$ nm) from the free DNA arms (yellow part of the complex, $h_{DNA} \sim 0.7$ nm) entering and exiting the complex.

Precise measurement of the length of each DNA fragment (respectively $L_+$ and $L_-$ for the longer and shorter arm) exiting the nucleosome have been performed. To measure each "arm" of the mono-nucleosome, the octamer part is excluded and the free DNA trajectory is obtained (Fig.1b) using morphological tools avoiding false skeletonization by heuristic algorithm ( cf *Material and Methods*). From the total DNA length that is un-wrapped around the histone octamer, we get the length of DNA organized by the histone octamer ($L_c = L_{tot} - L_+ - L_-$) as well as the nucleosome position with respect to the center of the sequence ($\Delta L = (L_+ - L_-)/2$).

The 2D histogram $L_c / \Delta L$ is plotted on Fig. 1c for 702 conventional short nucleosomes using a 2D sliding box as described in the *Material and Methods* section. The maximum of the 2D distribution is positioned at $L^* = 145$ bp and $\Delta L = 15$ bp, in qualitative agreement with the DNA template construction. The 2D mapping is an important tool to study nucleosome mobilization (see the SWI/SNF sliding section), since both variables are highly correlated during nucleosome sliding/remodeling. Quantitative information can be however also obtained by projecting such a 2D histogram on each axis. First, we have selected well positioned nucleosomes according to the expected position given by the DNA 601 template construction (0 bp < $\Delta L$ < 12 bp for short DNA fragments) and shown their DNA complexed length probability density function (red line, Fig. 1d). This distribution of the DNA length, organized by conventional octamer peaks at $L^* = 146 \pm 2$ bp, in quantitative agreement with the crystal structure of the nucleosome (26) and cryoEM measurements (25). The broadness of this distribution ($\sigma = 23$bp) might be explained by different nucleosomes wrapping conformations. We will explain later on, how this dispersion relates to DNA-histone interaction energies using a simple model.

We have used the same approach to study long nucleosomes (2D histogram not shown). Well positioned long nucleosomes according to the DNA sequence (12 bp < $\Delta L$ < 32 bp) have very similar probability distribution (blue line on Fig. 1d) than that obtained for short nucleosomes showing that the free linker DNA does not affect significantly the organization of complexed DNA for such nucleosomes.

We now select nucleosomes that have a complexed length in the range $L^* \pm \sigma_{Lc}$, where $\sigma_{Lc}$ is the standard deviation of the $L_c$ distribution, and their position distribution is displayed on Figure 1e. The peak values for each DNA fragment (9 ± 2 bp and 24 ± 2 bp for short and long nucleosomes respectively) is close to the expected value from the DNA template construct (2 bp and 22 bp for short and long DNA fragments respectively). Both distributions have a full width at half maximum that exceeds 20 bp. This width might arise from several features : asymmetric unwrapping of one of the two DNA arms, AFM uncertainty and dispersion in octamer position. However, it is not possible with these measurements to determine what is the contribution of each phenomenon. Next, we can see that the distribution width for longer fragments seems greater. After corrections of artifacts inherent to $L_+/L_-$ labeling (*cf* Supplemental Figure 2) these two position distributions are very similar showing that the free linker DNA does not affect either the DNA complexed length nor the positioning of such nucleosomes significantly.

We have shown in this section that AFM measurements give comparable estimations with other methods for both the positioning and the DNA wrapping of short 601 mononucleosomes. Furthermore, our experimental approach showed no difference in complexed length probability or nucleosome positioning dynamics for long and short DNA templates.

## For conventional and H2A.Bbd variant mononucleosomes

In order to investigate the influence of the octamer composition on the wrapping of DNA around the histone octamer, a H2A.Bbd histone variant was used instead of conventional H2A, in order to reconstitute mono-nucleosomes on a 255 bp DNA fragment. The H2A.Bbd variant nucleosomes were imaged by AFM (25) and using the same analysis as described above, only the well positioned nucleosomes ($\Delta L < 12$ bp) were selected. Their DNA complexed length distribution is plotted on Fig. 2a where it is compared to conventional mononucleosomes reconstituted on the same 601 positioning sequence, 255 bp long, with the same position range selection ($\Delta L < 12$ bp).

The average length of wrapped DNA is clearly different for the variant H2A.Bbd nucleosomes as the distribution peak value is $L^*_{H2A.Bbd} = 130 \pm 3$ bp instead of $L^*_{H2A} = 146 \pm 2$ bp for the conventional nucleosomes. Moreover the standard deviation of the distribution is clearly larger for the H2A.Bbd variant ($\sigma = 41$ bp to be compared to $\sigma = 23$ bp for the conventional nucleosomes). These differences show that the H2A.Bbd variant nucleosome is a more labile complex with less DNA wrapped around the octamer, in agreement with previous observations by AFM and cryo-EM (25). The difference in DNA complexed length suggests that ~10 bp at each end of nucleosomal DNA are released from the octamer. Therefore, AFM allows visualizing subtle differences in the nucleosome structure.

Finally, the DNA complexed length distribution is asymmetric for canonical nucleosomes. This asymmetry can be quantified by measuring their skewness $\tilde{\mu}_3$, defined as:

$$\tilde{\mu}_3 = \frac{\mu_3}{\mu_2^{3/2}} = \frac{\overline{(L_c - \overline{L_c})^3}}{\overline{((L_c - \overline{L_c})^2)}^{3/2}}$$

.We find $\tilde{\mu}_3 = -0.57 \pm 0.09$, the negative sign meaning that nucleosome conformations with sub-complexed DNA, as compared to the mean value 146 bp, are energetically more favorable than with over-complexed DNA. This can be interpreted within the simple model proposed below, based on relevant structural data information (26). Notice that for variant nucleosomes, the complexed length distribution is nearly symmetric ($\tilde{\mu}_3 \approx 0.01 \pm 0.16$), and this feature will also be discussed in the modeling section.

## Simple model of DNA complexed length distribution

It has been shown that 14 discrete contacts between DNA and histone octamer are responsible for the stability of the nucleosome (26).The energetic gain at these sites is made through electrostatic interactions and hydrogen bonding. At the length scale of the present analysis, the discreteness of binding sites is not relevant, and it will be replaced by a uniform effective adsorption energy $\varepsilon_{a<}$ per unit length, in units of kT/bp. The finite number of binding sites, or equivalently the finite DNA length $L^*$ complexed through these sites (146 bp for canonical nucleosomes, as determined both by the present experiments and crystal structure), is due to the specific locations of favorable interactions located at the surface of the histone octamer, forming a superhelical trajectory on which DNA is complexed. DNA wrapping around the histone core involves additional bending penalty characterized by the energy per unit length : $\varepsilon_b = \frac{kT \cdot L_p}{2R^2}$ where $L_p$ is the persistence length of DNA within classical linear elasticity and R the radius of the histone octamer. The stability of the nucleosome requires that the net energy per unit length is negative (energetic gain), and therefore : $\varepsilon_b < \varepsilon_{a<}$.

The experimental distributions of DNA complexed length show that more DNA can be wrapped around the octamer. For these additional base pairs, the net energy per unit length has to be positive, due mainly to bending cost. However, to allow for the possibility of some residual non specific (mainly electrostatic) attractive interactions beyond the 14 binding sites, the energetic gain of DNA contacting the octamer surface outside of the 14 sites superhelical path has a different value denoted $\varepsilon_{a>}$. The difference $\varepsilon_{a<} - \varepsilon_{a>}$ is then representative of the specificity of the 14 sites region.

Assuming that the energy reference is given by un-complexed straight DNA and octamer, the total energy for nucleosome is given by

$$\frac{E(L_c)}{kT} = \begin{cases} (\varepsilon_b - \varepsilon_{a<}) \cdot L_c & \text{if } L_c < L^* \text{ (sub-complexed nucleosome)} \\ (\varepsilon_b - \varepsilon_{a<}) \cdot L^* + (\varepsilon_b - \varepsilon_{a>}) \cdot (L_c - L^*) & \text{if } L_c > L^* \text{ (over-complexed nucleosome)} \end{cases} \quad (1)$$

The distribution of DNA complexed length is given by $P(L_c) \propto e^{-E(L_c)/(kT)}$. It is maximum for the characteristic length $L^*$, which characterizes the region of specific contacts. This length may vary for canonical and variant nucleosomes. The assumptions of energy linearity in wrapped DNA length and of the existence of $L^*$, lead to a double exponential distribution. By construction, one has the following constraints between effective energies $\varepsilon_{a>} < \varepsilon_b < \varepsilon_{a<}$.

It should be kept in mind that the effective values $\varepsilon_{a>}$, $\varepsilon_{a<}$ and $\varepsilon_b$ are representative of nucleosomes adsorbed on a charged flat surface. These values might differ for nucleosomes in bulk solution, as discussed below.

## *Extraction of the DNA complexed length parameters*

It is possible to extract some parameters from each distribution by using the physical model presented below, in order to interpret the experimental distribution of DNA complexed length. We found it more reliable to use global procedure for parameter determination, instead of fitting the multivariate distribution. Since we expect the DNA complexed length distribution to be described by a simple double-exponential model, the probability density function can be written as a skew-Laplace distribution which moments are calculated as :

$$P(L_c = L) = \frac{1}{2\sqrt{2}\sigma} \begin{cases} e^{-\frac{L-L^*}{\sqrt{2}(1-\varepsilon)\sigma}}, & \text{for } L > L^* \\ e^{+\frac{L-L^*}{\sqrt{2}(1+\varepsilon)\sigma}}, & \text{for } L < L^* \end{cases} \text{ and then } \begin{cases} \mu_1 = \overline{L_c} = L^* - 2\sqrt{2}\varepsilon\sigma \\ \mu_2 = \overline{(L_c - \overline{L_c})^2} = 4\sigma^2(1+\varepsilon^2) \\ \tilde{\mu}_3 = \frac{\overline{(L_c - \overline{L_c})^3}}{\mu_2^{3/2}} = \frac{4 - 50\sqrt{2}\varepsilon + 12\varepsilon^2 - 48\sqrt{2}\varepsilon^3}{4^{3/2}(1+\varepsilon^2)^{3/2}} \end{cases} \quad (2)$$

where $L^*$ is the most probable complexed length, $\varepsilon$ is the relative asymmetry of the skew-Laplace distribution and $\sigma$ is the mean decay length. The distribution normalization is taken on full real axis as a first approximation, thus neglecting finite size effects. Given the experimentally determined $\mu_1$, $\mu_2$ and $\mu_3$ parameters, we extract straightforwardly the parameter $L^*$, $\varepsilon$ and $\sigma$ by numerically solving the equation system (2).

Hence, we are able to measure *without any fitting* the parameters $L^*$, $\varepsilon$ and $\sigma$ by calculating the first three moments $\mu_1$, $\mu_2$ and $\mu_3$ of the DNA complexed length statistical series. In our case we thus have :

$$\begin{cases} \varepsilon_b - \varepsilon_{a<} = -\dfrac{1}{\sqrt{2}(1+\varepsilon)\sigma} = -\dfrac{1}{L_<} \\ \varepsilon_b - \varepsilon_{a>} = \dfrac{1}{\sqrt{2}(1-\varepsilon)\sigma} = \dfrac{1}{L_>} \end{cases} \text{ and then } E^{(specific)}_{ads} = \varepsilon_{a>} - \varepsilon_{a<} = \dfrac{\sqrt{2}}{(1-\varepsilon^2)\sigma}$$

To see the adequacy of this model with the experimental distribution, the function $P(L_c=L)$ is drawn for the parameters extracted from the experimental data using the same 20 bp-sliding box protocol as for the experimental complexed length distribution (Fig.2)

The results are summarized in table 1. The values of energies are expressed in units of kT per binding site, assuming 14 such sites along the 147 base pairs of DNA for canonical nucleosomes. Several comments are to be made on these values. First, the measured characteristic decay lengths corresponding to sub- ($L_<$) and over-complexed ($L_>$) DNA lengths (Table 1, (b) and (d)) are clearly higher than the intrinsic resolution of our AFM measurements (related to the tip size that correspond to ~ 10 bp, as checked by image analysis of the same naked DNA on the same surface and within the same experimental conditions - data not shown) for both conventional and variant nucleosomes, showing therefore the significance of the parameters extracted here. Hence, we are able to quantify the energetic of both sub- and over-complexed DNA length in a mono-nucleosome. For over-complexed DNA length, the energy has been converted artificially into units of kT per binding site for the sake of comparison, although the model assumes that there are no such binding sites beyond the 14 sites found in the crystal structure (26). If one assumes that over-complexed DNA length results solely from bending around the histone core ($\varepsilon_{a>} = 0$), the value found for $\varepsilon_b$ leads to a persistence length $L_p$ ~ 3.5 bp, a value definitely too small for double stranded DNA. Even more so, this energy is similar in amplitude to the energy of sub-complexed DNA length but with an opposite sign (Table 1, (c) and (e)). We conclude that it cannot simply be associated to a bending penalty, therefore justifying *a posteriori* the assumption of residual attractive interaction between DNA extra length and histone octamer.

The combination of experimental asymmetry of DNA complexed length distribution and the simple model allows quantifying the specificity of the 14 binding sites in the nucleosomes (Table 1, (f)). In particular this can be interpreted as a rough estimation of non-electrostatic contribution to adhesion energy between DNA and histone octamer.

*Comparison of model parameters extracted from data.*

These values have to be compared to other estimates reported in the literature. The net energetic gain per site can be compared to values extracted from experiments done in the group of J. Widom (68-70). The spirit of these experiments was to probe the transient exposure of DNA complexed length in a nucleosome by using different restriction enzymes acting at various well-defined sites along the DNA. The experimental results clearly demonstrate that DNA accessibility is strongly reduced when restriction sites are located far away from entry or exit of nucleosomal DNA, towards the dyad axis. From the experimental data, the authors extract a Boltzmann weight for different site exposures. This distribution should *a priori* be similar to the DNA complexed length distribution obtained in our work, except that only sub-complexed nucleosomes are probed. However, due to the use of different restriction enzymes with different sizes and mechanisms of action, there is an inherent uncertainty in the assignment of precise DNA complexed length with a free energy of the Boltzmann weight. In other words, only a range of energy per binding site can be extracted from these data. This has to be contrasted with most of previous works using Polach and Widom's data, which quote a *single* value of 2 kT per binding site (31). The range of net energetic gain we are able to estimate out of these data is between 0,5 to 3 kT per binding site. The value we extracted from our own measurements coincides therefore with the lower bound of this range. This might be due to the difference in the type of experiments used.

First, our observations are made on nucleosomes adsorbed on a charged substrate. This might change the energetics of nucleosome opening as compared to its value in solution. A theoretical estimation of this change is currently under progress (Castelnovo *et al*, work in preparation). Another significant difference between Polach and Widom's experiments and our work is the composition of the buffer, which is known to affect the nucleosome stability. In particular, the buffer used for restriction enzyme assays contains more magnesium ions (about 10 mM $MgCl_2$).

The specificity of DNA binding sites on histone octamer, as determined in Table 1 (e) can also be compared to values extracted from X-ray experiments performed in the group of T.J. Richmond (71). Indeed, by counting the hydrogen bonds per binding site found in this structure, one can estimate the specific contribution to the binding energy. These contributions range between 0.8 and 2 kT per binding site (72). Our estimate for conventional nucleosome falls in this range (1.1 kT per binding site).

Finally, the comparison between canonical and variant nucleosome shows that both the average complexed length and the energy per binding site are different. The most probable length $L^* = 127$ bp (Table 1 (0)) for the variant claims for either the absence or the strong weakening of at least 2 binding sites. Furthermore, the energy and therefore the stability of the nucleosome for the remaining binding sites is reduced ($\varepsilon_{H2A.Bbd} \sim 2/3\, \varepsilon_{H2A}$), in accordance with other experimental observations (16, 25, 73). We have shown in this section that a simple model using a linear energy for the DNA-histone interaction can be used to extract from the AFM data two important energetic parameters : the net energetic gain per site and the specific interaction between the DNA and the histone octamer per site. These values are in good agreement with previous biochemical and X-Ray studies done on conventional nucleosomes and for the first time are measured on a variant nucleosome.

# Visualization of nucleosome sliding and remodeling by SWI/SNF for conventional and variant nucleosomes.

After studying the nucleosomes in their equilibrium state, the same mononucleosomes were visualized in the presence of the SWI/SNF remodeling factor to validate the possibility for this direct imaging approach to acquire new information on the mechanism and dynamics of nucleosome sliding.

Centrally positioned conventional and variant mononucleosomes ($L_{tot} = 255$ bp) were incubated with SWI/SNF at 29°C in the presence or absence of ATP and then adsorbed on APTES-mica surfaces for AFM visualization. On figure 3, we report AFM images of mononucleosomes incubated without (Fig. 3a) and with (Fig. 3b) ATP for one hour. The sample containing no ATP is the control experiment to account for possible nucleosome thermally driven diffusion when incubated 1 h at 29 °C. The representative chosen set of AFM images of Fig.3 clearly shows that most of the nucleosomes are centered on the DNA template in the negative control (-ATP) whereas they rather exhibit end-position when SWI/SNF and ATP are present.

On AFM images, SWI/SNF motor is sometimes visible as a very large proteic complex, and if still attached to nucleosome prevents any image analysis of such objects. Our protocol does not include removing of SWI/SNF before deposition, even if by diluting the nucleosome/motor mix, one could expect detaching of some motors. Therefore, the motor per nucleosome ratio used in the sliding experiments is kept low with respect to biochemical assays (55) (roughly five time less SWI/SNF per nucleosome).

Using the same type of image analysis we were able to reconstruct 2D histograms $L_c/\Delta L$ (using a 2D sliding box as described in the Material and Methods section) at various time steps during nucleosome sliding : 0 (-ATP), 20 min and 1 hour (Fig. 4). We first notice that in the absence of ATP, SWI/SNF has apparently no effect on the $L_c/\Delta L$ map. The 2D distribution exhibits a single peak corresponding to the canonical nucleosome positioned as expected from the DNA template ($\alpha$ state). As a function of time in the presence of remodeling complex and ATP, new states appear : ($\beta$) corresponds to an over-complexed nucleosome having the same mean position $\Delta L$ value as ($\alpha$); this state could result from the capture of extra DNA (a loop of ~ 40 bp) inside the NCP induced by SWI/SNF. ($\beta$)-state is spread in the $\Delta L$ direction showing that this extra complexed DNA length (~ 40 bp) seems to exist for various positions of the nucleosome ($0 < \Delta L < 30$ bp). ($\gamma$) is the slided end-positioned nucleosome ($\Delta L \sim 50$ bp) having slightly less DNA wrapped around the histone octamer ($L_c \sim 125$ bp). The $\Delta L$ distance separating ($\alpha$) and ($\gamma$) states is close to the $L_c$ distance between ($\beta$) and ($\alpha$) states, meaning that the slided ($\gamma$)-state most likely results from the release of the ($\beta$)-state DNA loop (~ 40 bp). The fact that slided nucleosomes are sub-complexed i.e. their dyad has been moved beyond the expected end-position, has already been observed in other biochemical studies (74). Similarly, the anisotropic spreading of the ($\gamma$)-peak towards higher $\Delta L$ and lower $L_c$ is also consistent with this feature. We cannot exclude that a finite size effect of the DNA template could account for this feature. Finally, ($\delta$) is a wide state with a sub-complexed $L_c \sim 75$ bp, that could correspond to a tetrasome or hexasome. This state could be due to the loss of one wrapped DNA turn either from the $\alpha$ state or the ($\gamma$)-state. Nevertheless, one could notice that the ($\delta$)-state is missing on the '+ATP 20 min' map (Fig. 4b) where only few nucleosomes have been slided (weak $\gamma$ peak) whereas ($\delta$)-state nucleosomes are clearly visible on Fig. 4c ('+ATP 1 h'). This tends to show that ($\delta$)-state nucleosomes more likely arise from the loss of one DNA turn of the end-positioned nucleosomes (($\gamma$)-state).

We have seen that the 2D-mapping of nucleosome position and DNA complexed length allows characterizing the new states resulting from the ATP-dependent action of SWI/SNF on our 601 nucleosomes : an over-complexed state close to the 601 template center ($\beta$), a slided state ($\gamma$) and a sub-complexed state ($\delta$).

Again, more information can be gained by appropriate projections of these 2D-histograms. Nucleosomes having their DNA complexed length in the range $L^* \pm \sigma_{Lc}$ were selected ($L^*$ and $\sigma_{Lc}$ are respectively the maximum value and the standard deviation of the corresponding complexed length distribution) and their position distribution is plotted on Fig. 5a. For conventional nucleosomes with SWI/SNF but no ATP, the distributions obtained for nucleosome position (Fig. 5a) and DNA complexed length (Fig. 5b) are very similar to the case without any remodeling complex (Fig. 1b), showing no effect of thermally driven diffusion of mononucleosomes reconstituted on 601 positioning sequence in our conditions.

When incubation is increased in the presence of ATP (20 minutes and 1 hour), the position distribution of conventional nucleosomes is clearly changed (Fig. 5a). Indeed, as a function of incubation time, a second peak appears corresponding to the end-positioned nucleosomes ($\Delta L \sim 50$ bp, cf ($\gamma$)–state in Fig. 4c). After one hour of SWI/SNF action in presence of ATP the second peak height has increased at the expense of the primary peak. This corresponds to the situation were one third of the mono-nucleosomes are positioned at the end of the DNA template. It is interesting to note that during the remodeling factor action we do not see any significant increase in the amount of nucleosomes in an intermediate position (20 bp $< \Delta L <$ 40 bp). This provides experimental evidence that this remodeling factor moves centrally positioned nucleosomes directly to the end of our short DNA template.

Mainly, two situations can explain the bimodal position distribution of nucleosomes after the action of SWI/SNF. The first hypothesis is that the SWI/SNF complex is a processive molecular motor. As it will not detach from the nucleosome before it reaches the end of the DNA template, the elementary step of the SWI/SNF induced sliding might not be accessible. Indeed, in our experimental conditions, only nucleosomes without SWI/SNF complex attached can be analyzed. The other possibility is that SWI/SNF is weakly processive (SWI/SNF turnover rate is unknown) but with an elementary step of the order of 50 bp, which corresponds to the value measured by us and other approaches (75-77), and happens to be the length of free DNA arms in our case. Therefore, a single step would be enough for the motor to slide a nucleosome to an end-position and release the complex.

Nevertheless, another mechanism cannot be excluded by our data, where SWI/SNF action would consist of octamer destabilization followed by thermally driven diffusion towards the end-positioned entropically favored. In this situation, ATP-hydrolysis would only be involved in the nucleosome *'destabilization'* step.

In Fig. 5b, we show projections of the previous 2D-histograms along the DNA complexed length axis without any selection on their position. For conventional nucleosomes in the presence of SWI/SNF but no ATP, the complexed length distribution is similar to case with neither SWI/SNF nor ATP. However, the former distribution is larger due to the contribution of different nucleosome positioning. Then after 20 min, the distribution is broader (roughly twice) and shifted towards higher $L_c$. This might be attributed to the contributions of the different states ($\beta$, $\gamma$, $\delta$) identified in the Fig. 4b/c. The increase in $L_c$ mean value is likely due to the statistical weight of the over-complexed ($\beta$)-state.

The same sliding experiment was performed on H2A.Bbd variant nucleosomes in absence and in presence of ATP and analyzed through the projection of the 2D-histogram $L_c/\Delta L$. No significant effect of SWI/SNF complexes in presence of ATP on the position distribution of H2A.Bbd variant nucleosomes is observed (Fig. 5c). This corroborates previous findings using biochemical sliding assay done on 5S and 601 positioning sequence (55). However in AFM measurements, the full position distribution is accessed directly with a resolution better than 10 bp (the size of AFM tip). This variant nucleosome sliding assay shows the reproducibility of our experimental approach as not only the position distribution mean value is constant during one hour in the presence of SWI/SNF and ATP, but also the complete position distribution remains constant (Fig. 5c). Similarly, SWI/SNF in presence of ATP does not seem to influence the DNA complexed length distribution of H2A.Bbd nucleosomes (Fig. 5d).

# CONCLUSION

In summary, we have shown that AFM combined with a systematic computer analysis is a powerful tool to determine the structure of conventional and variant mononucleosomes at equilibrium and after the action of ATP-dependent cellular machineries. With this technique we have quantified simultaneously two important and closely coupled variables : the DNA complexed length and the position of mono-nucleosomes along the 601 DNA template. For each of these two distributions, the most probable value is in perfect agreement with measurements done by other methods that give access to one of these two parameters only. In addition, to explain the experimental complexed length distribution, we have developed a simple model that uses the experimental shape of DNA complexed length distributions to quantify the interaction of DNA with histones. With this model, we extract both the net energetic gain for sub-complexed nucleosomes and the estimation of the non-electrostatic contribution to the adhesion energy between DNA and histone octamer .

We further show that H2A.Bbd variant and conventional nucleosomes exhibit clear differences in DNA complexed length and in their ability to be slided by SWI/SNF. Indeed, these variant nucleosomes organize less DNA on average than conventional nucleosomes, and present larger opening and closing fluctuations. Moreover, the whole position distribution as well as complexed length distribution remain unchanged showing H2A.Bbd variant is neither displaced nor remodeled by SWI/SNF complex.

Finally, we have plotted $L_c/\Delta L$ as a 2D map of the nucleosome states. This representation is well suited to highlight the various nucleosome states that appear during the SWI/SNF action. For example, as a function of time, we have evidenced the formation of an over-complexed state followed by the appearance of a slided state. More quantitative information can be obtained by appropriate projections of the 2D-histograms, as for instance the bimodal position distribution induced by SWI/SNF sliding on conventional nucleosomes, suggesting two possible scenarii : a processive action of the molecular motor (no intermediate position visualized) or an elementary stepping length (~ 40 bp) of the size of the free DNA arms (~ 50 bp). The short length of DNA templates and lack of directionality in our position analysis prevent us from discriminating between these two hypotheses, and further experiments on long oriented mononucleosomes are needed to get more insights into the molecular mechanism of SWI/SNF action. The present results as well as preliminary data on longer oriented templates prove nevertheless that this extension will provide useful information on remodeling mechanisms of SWI/SNF.

A further perspective of this AFM study will be to test the effect of the flanking DNA sequences on the conformation and dynamics of 601 nucleosomes. Nevertheless, in order to test sequence effect, nucleosomes should be reconstituted on less positioning sequences (5S rDNA for example) or non-positioning sequences, but this will complicate significantly the nucleosome sliding analysis as the initial position distribution of the nucleosome is expected to be broader in this case.

# ACKNOWLEDGMENTS

We thank Dimitar Angelov and Hervé Ménoni for various forms of help with nucleosome reconstitution and sliding assays and Cécile-Marie Doyen for producing H2A.Bbd variant histones. We are grateful to Stefan Dimitrov, Dimitar Angelov, Françoise Argoul, Alain Arneodo, Cédric Vaillant and Phillipe Bouvet for fruitful discussions. We thank

Ali Hamiche for providing us with the ySWI/SNF complex. P.St-J. acknowledges CRSNG for financial support. This work was supported by the CPER 'Nouvelles Approches Physiques des Sciences du Vivant'.

## REFERENCES

1. van Holde, K. 1988. Chromatin. Springer-Verlag KG, Berlin.
2. Beato, M., and K. Eisfeld. 1997. Transcription factor access to chromatin. Nucleic acids research 25:3559-3563.
3. Becker, P. B. 2002. Nucleosome sliding: facts and fiction. The EMBO journal 21:4749-4753.
4. Henikoff, S., and K. Ahmad. 2005. Assembly of variant histones into chromatin. Annual review of cell and developmental biology 21:133-153.
5. Henikoff, S., T. Furuyama, and K. Ahmad. 2004. Histone variants, nucleosome assembly and epigenetic inheritance. Trends Genet 20:320-326.
6. Strahl, B. D., and C. D. Allis. 2000. The language of covalent histone modifications. Nature 403:41-45.
7. Peterson, C. L., and J. L. Workman. 2000. Promoter targeting and chromatin remodeling by the SWI/SNF complex. Current opinion in genetics & development 10:187-192.
8. Travers, A. 1999. An engine for nucleosome remodeling. Cell 96:311-314.
9. Tsukiyama, T., and C. Wu. 1997. Chromatin remodeling and transcription. Current opinion in genetics & development 7:182-191.
10. Hamiche, A., J. G. Kang, C. Dennis, H. Xiao, and C. Wu. 2001. Histone tails modulate nucleosome mobility and regulate ATP-dependent nucleosome sliding by NURF. Proceedings of the National Academy of Sciences of the United States of America 98:14316-14321.
11. Langst, G., E. J. Bonte, D. F. Corona, and P. B. Becker. 1999. Nucleosome movement by CHRAC and ISWI without disruption or trans-displacement of the histone octamer. Cell 97:843-852.
12. Lorch, Y., M. Zhang, and R. D. Kornberg. 1999. Histone octamer transfer by a chromatin-remodeling complex. Cell 96:389-392.
13. Whitehouse, I., A. Flaus, B. R. Cairns, M. F. White, J. L. Workman, and T. Owen-Hughes. 1999. Nucleosome mobilization catalysed by the yeast SWI/SNF complex. Nature 400:784-787.
14. Abbott, D. W., V. S. Ivanova, X. Wang, W. M. Bonner, and J. Ausio. 2001. Characterization of the stability and folding of H2A.Z chromatin particles: implications for transcriptional activation. The Journal of biological chemistry 276:41945-41949.
15. Angelov, D., A. Molla, P. Y. Perche, F. Hans, J. Cote, S. Khochbin, P. Bouvet, and S. Dimitrov. 2003. The histone variant macroH2A interferes with transcription factor binding and SWI/SNF nucleosome remodeling. Molecular cell 11:1033-1041.
16. Bao, Y., K. Konesky, Y. J. Park, S. Rosu, P. N. Dyer, D. Rangasamy, D. J. Tremethick, P. J. Laybourn, and K. Luger. 2004. Nucleosomes containing the histone variant H2A.Bbd organize only 118 base pairs of DNA. The EMBO journal 23:3314-3324.
17. Chakravarthy, S., Y. Bao, V. A. Roberts, D. Tremethick, and K. Luger. 2004. Structural characterization of histone H2A variants. Cold Spring Harbor symposia on quantitative biology 69:227-234.

18. Fan, J. Y., D. Rangasamy, K. Luger, and D. J. Tremethick. 2004. H2A.Z alters the nucleosome surface to promote HP1alpha-mediated chromatin fiber folding. Molecular cell 16:655-661.
19. Suto, R. K., M. J. Clarkson, D. J. Tremethick, and K. Luger. 2000. Crystal structure of a nucleosome core particle containing the variant histone H2A.Z. Nature structural biology 7:1121-1124.
20. Ahmad, K., and S. Henikoff. 2002. Epigenetic consequences of nucleosome dynamics. Cell 111:281-284.
21. Ausio, J., and D. W. Abbott. 2002. The many tales of a tail: carboxyl-terminal tail heterogeneity specializes histone H2A variants for defined chromatin function. Biochemistry 41:5945-5949.
22. Kamakaka, R. T., and S. Biggins. 2005. Histone variants: deviants? Genes & development 19:295-310.
23. Sarma, K., and D. Reinberg. 2005. Histone variants meet their match. Nature reviews 6:139-149.
24. Chadwick, B. P., and H. F. Willard. 2001. A novel chromatin protein, distantly related to histone H2A, is largely excluded from the inactive X chromosome. The Journal of cell biology 152:375-384.
25. Doyen, C. M., F. Montel, T. Gautier, H. Menoni, C. Claudet, M. Delacour-Larose, D. Angelov, A. Hamiche, J. Bednar, C. Faivre-Moskalenko, P. Bouvet, and S. Dimitrov. 2006. Dissection of the unusual structural and functional properties of the variant H2A.Bbd nucleosome. The EMBO journal 25:4234-4244.
26. Luger, K., A. W. Mader, R. K. Richmond, D. F. Sargent, and T. J. Richmond. 1997. Crystal structure of the nucleosome core particle at 2.8 A resolution. Nature 389:251-260.
27. Polach, K. J., and J. Widom. 1999. Restriction enzymes as probes of nucleosome stability and dynamics. Methods in enzymology 304:278-298.
28. Wu, C., and A. Travers. 2004. A 'one-pot' assay for the accessibility of DNA in a nucleosome core particle. Nucleic acids research 32:e122.
29. Tomschik, M., H. Zheng, K. van Holde, J. Zlatanova, and S. H. Leuba. 2005. Fast, long-range, reversible conformational fluctuations in nucleosomes revealed by single-pair fluorescence resonance energy transfer. Proceedings of the National Academy of Sciences of the United States of America 102:3278-3283.
30. Toth, K., N. Brun, and J. Langowski. 2001. Trajectory of nucleosomal linker DNA studied by fluorescence resonance energy transfer. Biochemistry 40:6921-6928.
31. Schiessel, H. 2003. The physics of chromatin. R699-R774.
32. Bishop, T. C. 2005. Molecular dynamics simulations of a nucleosome and free DNA. Journal of biomolecular structure & dynamics 22:673-686.
33. Sharma, S., F. Ding, and N. V. Dokholyan. 2007. Multiscale modeling of nucleosome dynamics. Biophysical journal 92:1457-1470.
34. Wedemann, G., and J. Langowski. 2002. Computer simulation of the 30-nanometer chromatin fiber. Biophysical journal 82:2847-2859.
35. Ruscio, J. Z., and A. Onufriev. 2006. A computational study of nucleosomal DNA flexibility. Biophysical journal 91:4121-4132.
36. Binnig, G., C. F. Quate, and C. Gerber. 1986. Atomic force microscope. Physical Review Letters 56:930-933.
37. Zlatanova, J., and S. H. Leuba. 2003. Chromatin fibers, one-at-a-time. Journal of molecular biology 331:1-19.

38. Ji, M., P. Hou, Z. Lu, and N. He. 2004. Covalent immobilization of DNA onto functionalized mica for atomic force microscopy. Journal of nanoscience and nanotechnology 4:580-584.
39. Lyubchenko, Y., L. Shlyakhtenko, R. Harrington, P. Oden, and S. Lindsay. 1993. Atomic force microscopy of long DNA: imaging in air and under water. Proceedings of the National Academy of Sciences of the United States of America 90:2137-2140.
40. Lyubchenko, Y. L., P. I. Oden, D. Lampner, S. M. Lindsay, and K. A. Dunker. 1993. Atomic force microscopy of DNA and bacteriophage in air, water and propanol: the role of adhesion forces. Nucleic acids research 21:1117-1123.
41. Podesta, A., L. Imperadori, W. Colnaghi, L. Finzi, P. Milani, and D. Dunlap. 2004. Atomic force microscopy study of DNA deposited on poly L-ornithine-coated mica. Journal of microscopy 215:236-240.
42. Podesta, A., M. Indrieri, D. Brogioli, G. S. Manning, P. Milani, R. Guerra, L. Finzi, and D. Dunlap. 2005. Positively charged surfaces increase the flexibility of DNA. Biophysical journal 89:2558-2563.
43. Umemura, K., J. Komatsu, T. Uchihashi, N. Choi, S. Ikawa, T. Nishinaka, T. Shibata, Y. Nakayama, S. Katsura, A. Mizuno, H. Tokumoto, M. Ishikawa, and R. Kuroda. 2001. Atomic force microscopy of RecA--DNA complexes using a carbon nanotube tip. Biochemical and biophysical research communications 281:390-395.
44. Wang, H., R. Bash, J. G. Yodh, G. L. Hager, D. Lohr, and S. M. Lindsay. 2002. Glutaraldehyde modified mica: a new surface for atomic force microscopy of chromatin. Biophysical journal 83:3619-3625.
45. Kepert, J. F., J. Mazurkiewicz, G. L. Heuvelman, K. F. Toth, and K. Rippe. 2005. NAP1 modulates binding of linker histone H1 to chromatin and induces an extended chromatin fiber conformation. The Journal of biological chemistry 280:34063-34072.
46. Kepert, J. F., K. F. Toth, M. Caudron, N. Mucke, J. Langowski, and K. Rippe. 2003. Conformation of reconstituted mononucleosomes and effect of linker histone H1 binding studied by scanning force microscopy. Biophysical journal 85:4012-4022.
47. Rivetti, C., M. Guthold, and C. Bustamante. 1999. Wrapping of DNA around the E.coli RNA polymerase open promoter complex. The EMBO journal 18:4464-4475.
48. Aoyagi, S., P. A. Wade, and J. J. Hayes. 2003. Nucleosome sliding induced by the xMi-2 complex does not occur exclusively via a simple twist-diffusion mechanism. The Journal of biological chemistry 278:30562-30568.
49. Johnson, C. N., N. L. Adkins, and P. Georgel. 2005. Chromatin remodeling complexes: ATP-dependent machines in action. Biochemistry and cell biology = Biochimie et biologie cellulaire 83:405-417.
50. O'Donohue, M. F., I. Duband-Goulet, A. Hamiche, and A. Prunell. 1994. Octamer displacement and redistribution in transcription of single nucleosomes. Nucleic acids research 22:937-945.
51. Furrer, P., J. Bednar, J. Dubochet, A. Hamiche, and A. Prunell. 1995. DNA at the entry-exit of the nucleosome observed by cryoelectron microscopy. Journal of structural biology 114:177-183.
52. Li, G., M. Levitus, C. Bustamante, and J. Widom. 2005. Rapid spontaneous accessibility of nucleosomal DNA. Nature structural & molecular biology 12:46-53.
53. Lowary, P. T., and J. Widom. 1998. New DNA sequence rules for high affinity binding to histone octamer and sequence-directed nucleosome positioning. Journal of molecular biology 276:19-42.
54. Luger, K., T. J. Rechsteiner, and T. J. Richmond. 1999. Expression and purification of recombinant histones and nucleosome reconstitution. Methods in molecular biology (Clifton, N.J 119:1-16.

55. Angelov, D., A. Verdel, W. An, V. Bondarenko, F. Hans, C. M. Doyen, V. M. Studitsky, A. Hamiche, R. G. Roeder, P. Bouvet, and S. Dimitrov. 2004. SWI/SNF remodeling and p300-dependent transcription of histone variant H2ABbd nucleosomal arrays. The EMBO journal 23:3815-3824.
56. Cote, J., J. Quinn, J. L. Workman, and C. L. Peterson. 1994. Stimulation of GAL4 derivative binding to nucleosomal DNA by the yeast SWI/SNF complex. Science 265:53-60.
57. Mutskov, V., D. Gerber, D. Angelov, J. Ausio, J. Workman, and S. Dimitrov. 1998. Persistent interactions of core histone tails with nucleosomal DNA following acetylation and transcription factor binding. Molecular and cellular biology 18:6293-6304.
58. Rivetti, C., M. Guthold, and C. Bustamante. 1996. Scanning force microscopy of DNA deposited onto mica: equilibration versus kinetic trapping studied by statistical polymer chain analysis. Journal of molecular biology 264:919-932.
59. Valle, F., M. Favre, P. De Los Rios, A. Rosa, and G. Dietler. 2005. Scaling exponents and probability distributions of DNA end-to-end distance. Phys Rev Lett 95:158105.
60. Ye, J. Y., K. Umemura, M. Ishikawa, and R. Kuroda. 2000. Atomic force microscopy of DNA molecules stretched by spin-coating technique. Analytical biochemistry 281:21-25.
61. Gonzalez, and Woods. 2002. Digital Image Processing 2nd Edition. Prentice Hall, New York.
62. Rigotti, D. J., B. Kokona, T. Horne, E. K. Acton, C. D. Lederman, K. A. Johnson, R. S. Manning, S. A. Kane, W. F. Smith, and R. Fairman. 2005. Quantitative atomic force microscopy image analysis of unusual filaments formed by the Acanthamoeba castellanii myosin II rod domain. Analytical biochemistry 346:189-200.
63. Rivetti, C., and S. Codeluppi. 2001. Accurate length determination of DNA molecules visualized by atomic force microscopy: evidence for a partial B- to A-form transition on mica. Ultramicroscopy 87:55-66.
64. Spisz, T. S., Y. Fang, R. H. Reeves, C. K. Seymour, I. N. Bankman, and J. H. Hoh. 1998. Automated sizing of DNA fragments in atomic force microscope images. Medical & biological engineering & computing 36:667-672.
65. Canny, J. 1986. A computational approach to edge detection. IEEE Computer Society Washington, DC, USA. 679-698.
66. Varadhan, G., W. Robinett, D. Erie, and R. M. T. Ii. 2003. Fast Simulation of Atomic-Force-Microscope Imaging of Atomic and Polygonal Surfaces Using Graphics Hardware. SPIE. 116.
67. Villarrubia, J. S. 1997. Algorithms for scanned probe microscope image simulation, surface reconstruction, and tip estimation. 425-453.
68. Polach, K. J., and J. Widom. 1996. A model for the cooperative binding of eukaryotic regulatory proteins to nucleosomal target sites. Journal of molecular biology 258:800-812.
69. Polach, K. J., and J. Widom. 1995. Mechanism of protein access to specific DNA sequences in chromatin: a dynamic equilibrium model for gene regulation. Journal of molecular biology 254:130-149.
70. Anderson, J. D., and J. Widom. 2000. Sequence and position-dependence of the equilibrium accessibility of nucleosomal DNA target sites. Journal of molecular biology 296:979-987.
71. Davey, C. A., D. F. Sargent, K. Luger, A. W. Maeder, and T. J. Richmond. 2002. Solvent mediated interactions in the structure of the nucleosome core particle at 1.9 a resolution. Journal of molecular biology 319:1097-1113.

72. Nikova, D. N., L. H. Pope, M. L. Bennink, K. A. van Leijenhorst-Groener, K. van der Werf, and J. Greve. 2004. Unexpected binding motifs for subnucleosomal particles revealed by atomic force microscopy. Biophysical journal 87:4135-4145.
73. Gautier, T., D. W. Abbott, A. Molla, A. Verdel, J. Ausio, and S. Dimitrov. 2004. Histone variant H2ABbd confers lower stability to the nucleosome. EMBO reports 5:715-720.
74. Kassabov, S. R., N. M. Henry, M. Zofall, T. Tsukiyama, and B. Bartholomew. 2002. High-resolution mapping of changes in histone-DNA contacts of nucleosomes remodeled by ISW2. Molecular and cellular biology 22:7524-7534.
75. Shundrovsky, A., C. L. Smith, J. T. Lis, C. L. Peterson, and M. D. Wang. 2006. Probing SWI/SNF remodeling of the nucleosome by unzipping single DNA molecules. Nature structural & molecular biology 13:549-554.
76. Zhang, Y., C. L. Smith, A. Saha, S. W. Grill, S. Mihardja, S. B. Smith, B. R. Cairns, C. L. Peterson, and C. Bustamante. 2006. DNA Translocation and Loop Formation Mechanism of Chromatin Remodeling by SWI/SNF and RSC. Molecular cell 24:559-568.
77. Zofall, M., J. Persinger, S. R. Kassabov, and B. Bartholomew. 2006. Chromatin remodeling by ISW2 and SWI/SNF requires DNA translocation inside the nucleosome. Nature structural & molecular biology 13:339-346.

## TABLE 1

|  | (a) $\langle L_c \rangle$ (bp) | (b) decay length $L_<$ (bp) | (c) $\varepsilon_b - \varepsilon_{a<}$ (kT per site) | (d) decay length $L_>$ (bp) | (e) $\varepsilon_b - \varepsilon_{a>}$ (kT per site) | (f) $\varepsilon_{a<} - \varepsilon_{a>}$ (kT per site) |
|---|---|---|---|---|---|---|
| conventional nucleosome | 146 ± 2 | 22 ± 1.6 | -0.479 ± 0.045 | 17 ± 1.4 | 0.61 ± 0.064 | 1.1 ± 0.072 |
| variant nucleosome | 127 ± 3 | 31 ± 1.5 | -0.33 ± 0.022 | 27 ± 1.5 | 0.39 ± 0.026 | 0.72 ± 0.021 |
| site exposure model (g) |  |  | -3 <...< -0.5 |  |  |  |
| crystal structure (h) | 147 |  |  |  |  | 0.8<···<2 |

**Caption Table1** : Summary of model parameters extracted from experimental data as explained in *Materials and methods*. All energies are expressed in units of kT per binding site. DNA lengths are expressed in bp. (a) Average complexed length (b) Characteristic length $L_<$ of exponential decay towards sub-complexed DNA length. (c) Energy per binding site ($1/L_<$) for sub-complexed DNA length. (d) Characteristic length $L_>$ of exponential decay, towards over-complexed DNA length. (e) Energy per binding site ($1/L_>$) for over-complexed DNA length. (f) Asymmetry of adhesion energy per binding site between sub- and over-complexed DNA length. (g) Range of values extracted from Polach and Widom (27, 69) data using the site exposure model. (h) Range of values extracted from Davey and Richmond (71) data using X-ray crystal structure of the nuclear core particle. Uncertainty values are determined using the central limit theorem and a propagation of uncertainty calculus detailed in supplemental data. N(H2A conventional) = 301 nucleosomes. N(H2A.Bbd variant) = 252 nucleosomes.

# FIGURE CAPTIONS

**Figure 1 : AFM visualization of centered mononucleosomes with short and long arms.** **(a)** AFM topography image of mono-nucleosomes reconstituted on 356 bp 601 positioning sequence. Color scale : from 0 to 1.5 nm. X/Y scale bar : 100 nm. **(b)** Zoom in the AFM topography image of a centered mono-nucleosome and the result of the image analysis. Black line : contour of the mono-nucleosome. Blue point : centroid of the histone octamer. Blue dot circle : excluded area of the histone octamer. Blue line : skeletons of the free DNA arms. Color scale : from 0 to 1.5 nm. X/Y Scale bar : 20 nm. The longest arm is named $L_+$ and the shortest $L_-$. DNA complexed length is deduced by $L_c = L_{tot} - L_- - L_+$ where $L_{tot}$ is in this case 356 bp. The position of the nucleosome relatively to the center of the sequence is calculated by $\Delta L = (L_+ - L_-)/2$. **(c)** 2D histogram $L_c/\Delta L$ representing the DNA complexed length $L_c$ along with the nucleosome position $\Delta L$ for a short DNA fragment of 255 bp (N = 702 nucleosomes). **(d)** Probability density function of the DNA complexed length $L_c$ for a short DNA fragment (255 bp, purple line) and for a long DNA fragment (356 bp, blue line) obtained by selecting the well positioned nucleosomes ($0 < \Delta L < 12$ bp and $12 < \Delta L < 32$ bp for the short and long fragments respectively) and projecting the 2D map along the y-axis. **(e)** Probability density function of the $\Delta L$ nucleosome position for a short DNA fragment (255 bp, purple line) and for a long DNA fragment (356 bp, blue line) obtained by selecting nucleosomes having their DNA complexed length $L_c$ in the range 123 bp < $L_c$ < 169 bp for both fragments, and projecting the 2D map along the x-axis.

**Figure 2 : AFM Visualization of centered H2A.Bbd variant and H2A conventional mononucleosome.**

**(a)** Probability density function of the DNA complexed length $L_c$ for a short DNA fragment (255 bp) with conventional H2A (solid thick line) and with variant H2A.Bbd (dotted thick line) nucleosome. Simple model for conventional and variant nucleosomes (respectively solid and dashed thin lines). **(b)** Description of the model used to measure the DNA-histone adsorption energies per bp ($\varepsilon_{a<}$ and $\varepsilon_{a>}$) and the DNA bending energy per bp ($\varepsilon_b$) (dotted line). Representation of the model using the 20bp-sliding-box procedure (dotted dashed line). L* corresponds to the most probable DNA complexed length of the distribution.

**Figure 3 : AFM Visualization of the sliding of centered mononucleosomes by the remodeling complex SWI/SNF**.

AFM topography image of mononucleosomes reconstituted on 255 bp 601 positioning sequence, incubated at 29°C with SWI/SNF for one hour **(a)** in the absence and **(b)** in the presence of ATP. Color scale : from 0 to 1.5 nm. X/Y Scale bar : 150 nm. Zoom in the AFM topography image of a **(c)** centered mononucleosome and **(d)** end-positioned mononucleosome the result of the image analysis. Black line : contour of the mono-nucleosome. Blue point : centroid of the histone octamer. Blue line : skeletons of the free DNA arms. Color scale : from 0 to 1.5 nm. X/Y Scale bar : 20 nm.

**Figure 4 : Evolution of nucleosome $L_c/\Delta L$ map during nucleosome sliding by SWI/SNF complex for conventional nucleosome.**

2D histogram $L_c/\Delta L$ representing the DNA complexed length $L_c$ along with the nucleosome position $\Delta L$ for a conventional nucleosome reconstituted on a short DNA fragment (255 bp) in the presence of remodeling complex SWI/SNF **(a)** without ATP (1h at 29°C) **(b)** with ATP (20 min at 29°C) and **(c)** with ATP (1h at 29°C). **(d)** Representation of the nucleosome $L_c/\Delta L$ states for (α), (β), (χ) and (δ) positions as pointed on the 2D maps. N(-ATP, 1h at 29°C) = 692 nucleosomes, N(+ATP, 20 min at 29°C) = 245 nucleosomes, N(+ATP, 1h at 29°C) = 655 nucleosomes.

**Figure 5 : Evolution of nucleosome position and DNA complexed length distributions during nucleosome sliding by SWI/SNF complex, for conventional and variant nucleosome.**

Nucleosome position $\Delta L$ **(a)** and DNA complexed length $L_c$ **(b)** distributions as a function of time (0, 20 min, 1 hour) in the presence of SWI/SNF, for conventional mono-nucleosomes reconstituted on 255 bp long 601 positioning sequence. Nucleosome position $\Delta L$ **(c)** and DNA complexed length $L_c$ **(d)** distributions as a function of time (0, 1 hour) in the presence of SWI/SNF, for H2A.Bbd variant mono-nucleosomes reconstituted on the same DNA template. The zero time is given by the control in the absence of ATP (solid purple line). For each $\Delta L$ position distribution, only nucleosomes having their complexed length in the range $L_c^* \pm \sigma_{Lc}$ are selected. For the sake of figure clarity, error bars are only depicted on 2 distributions of graph (a).

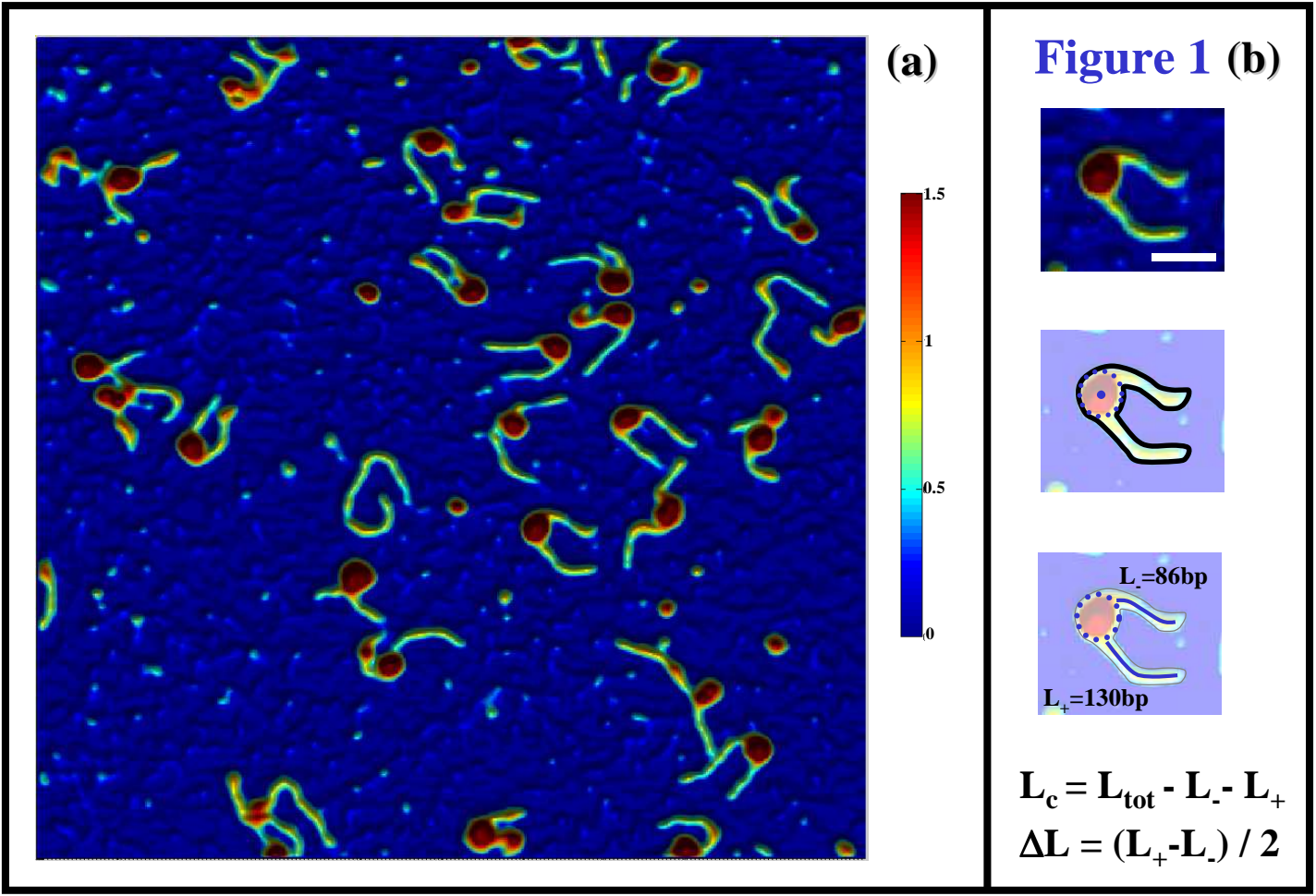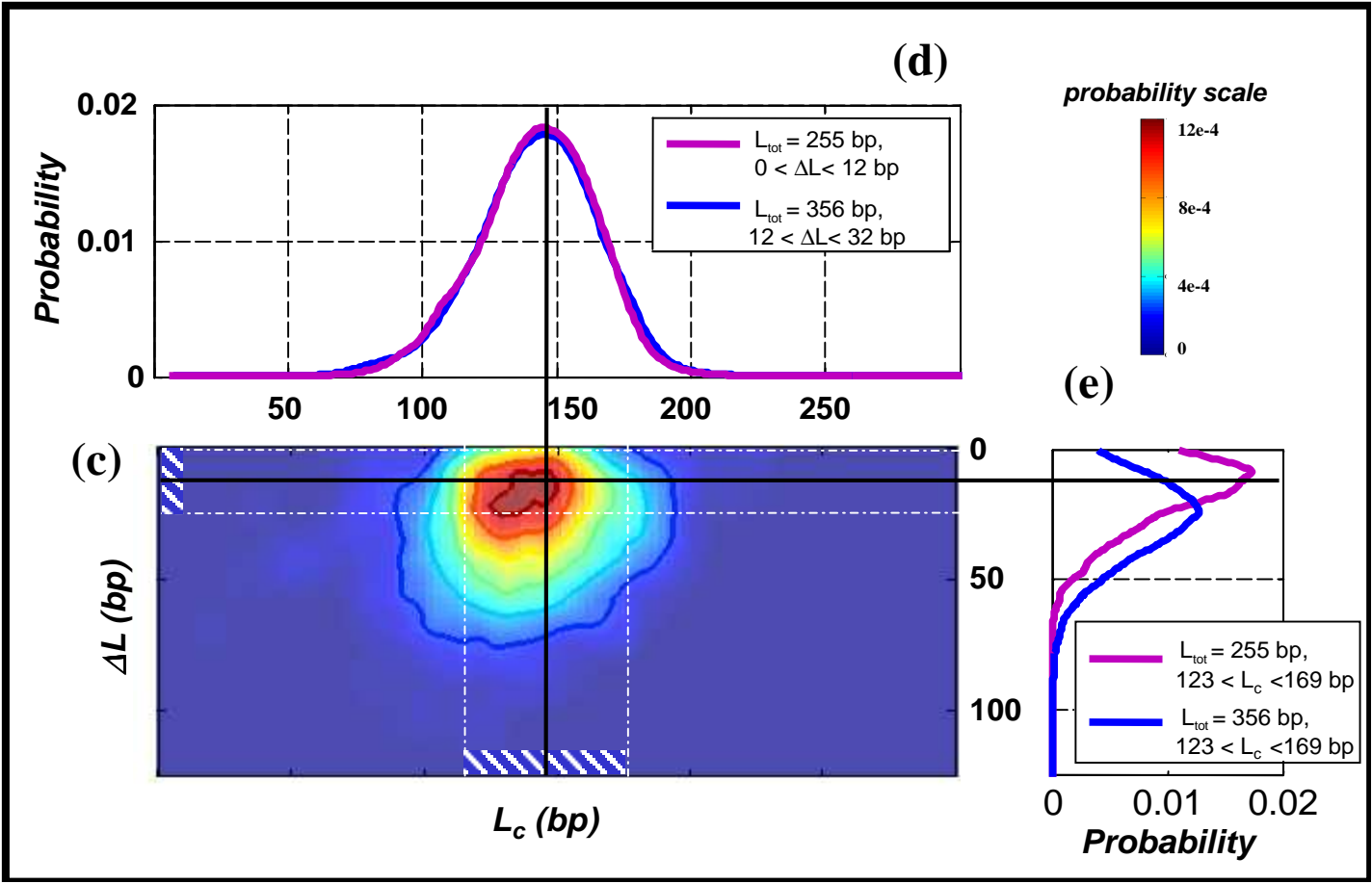

Figure 1

**Figure 2**

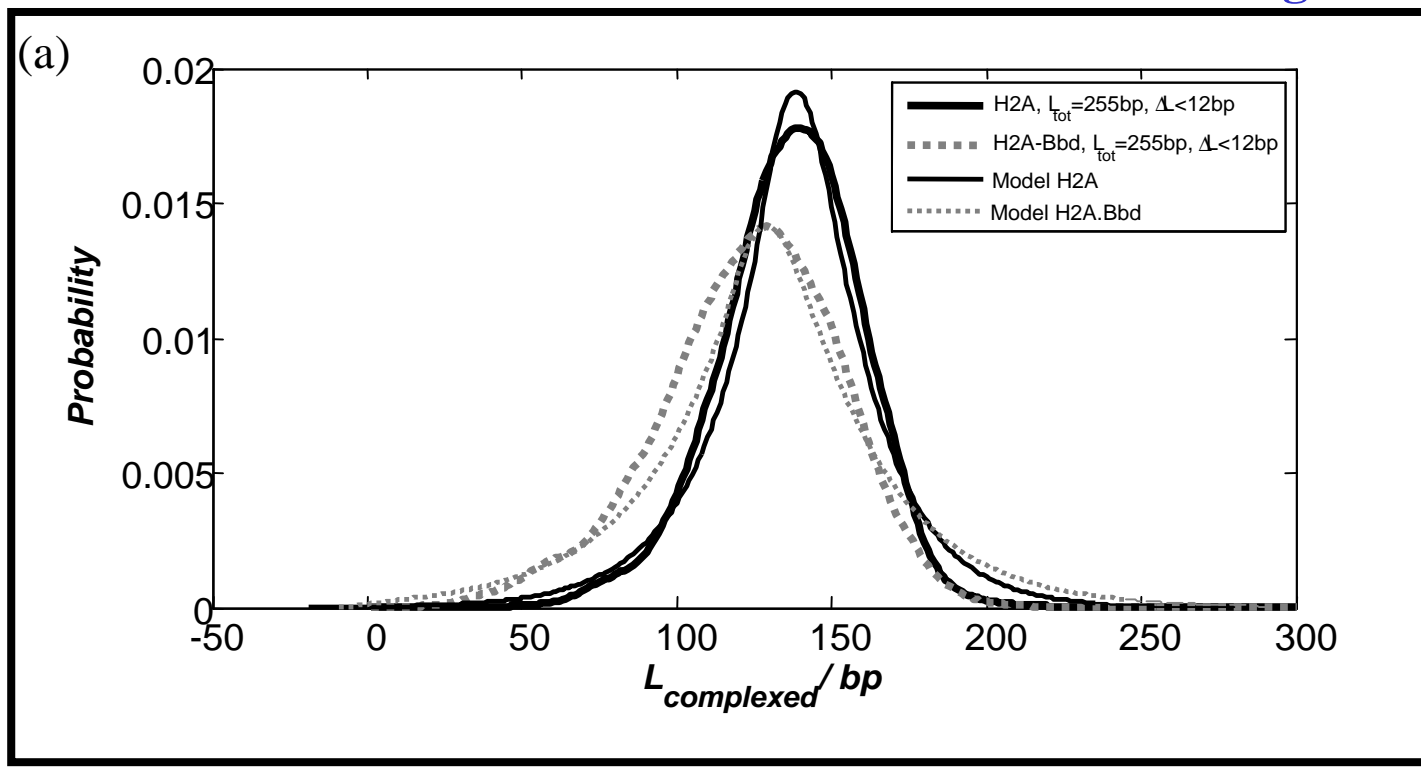

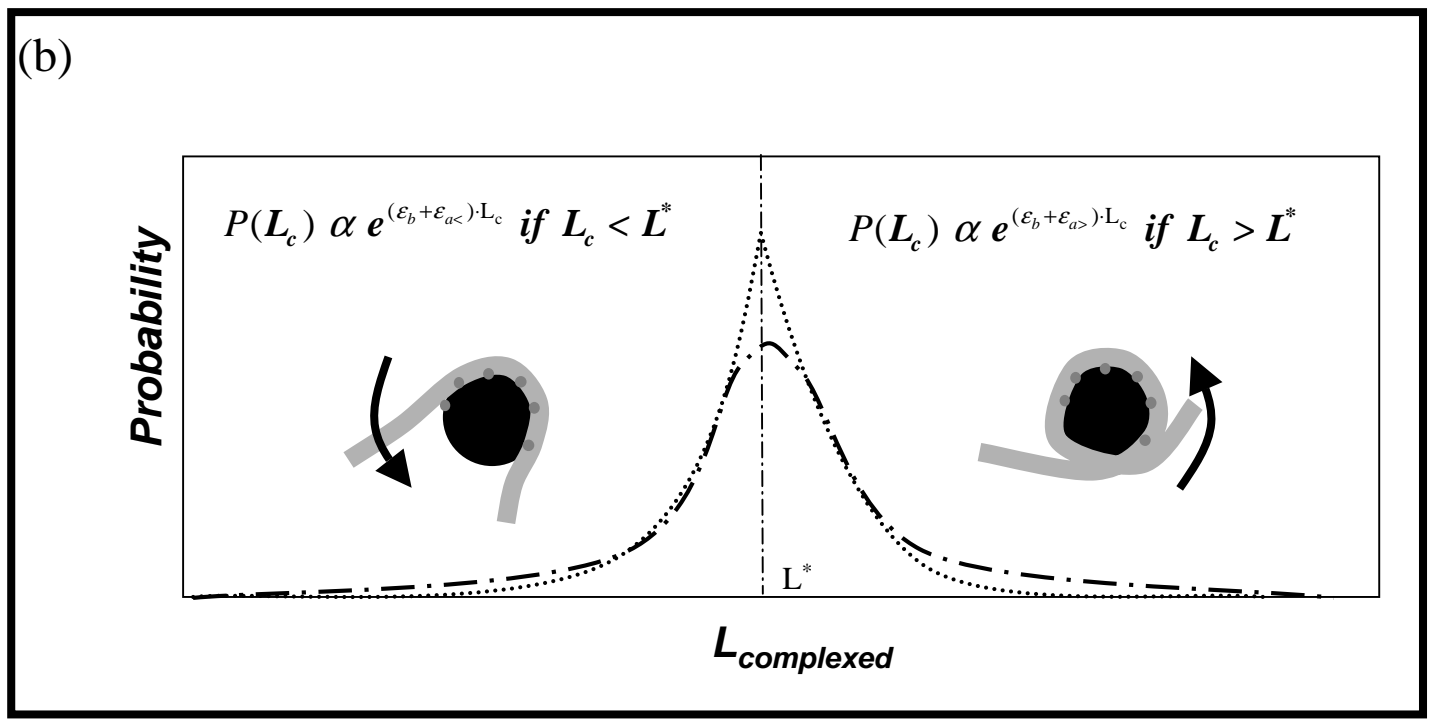

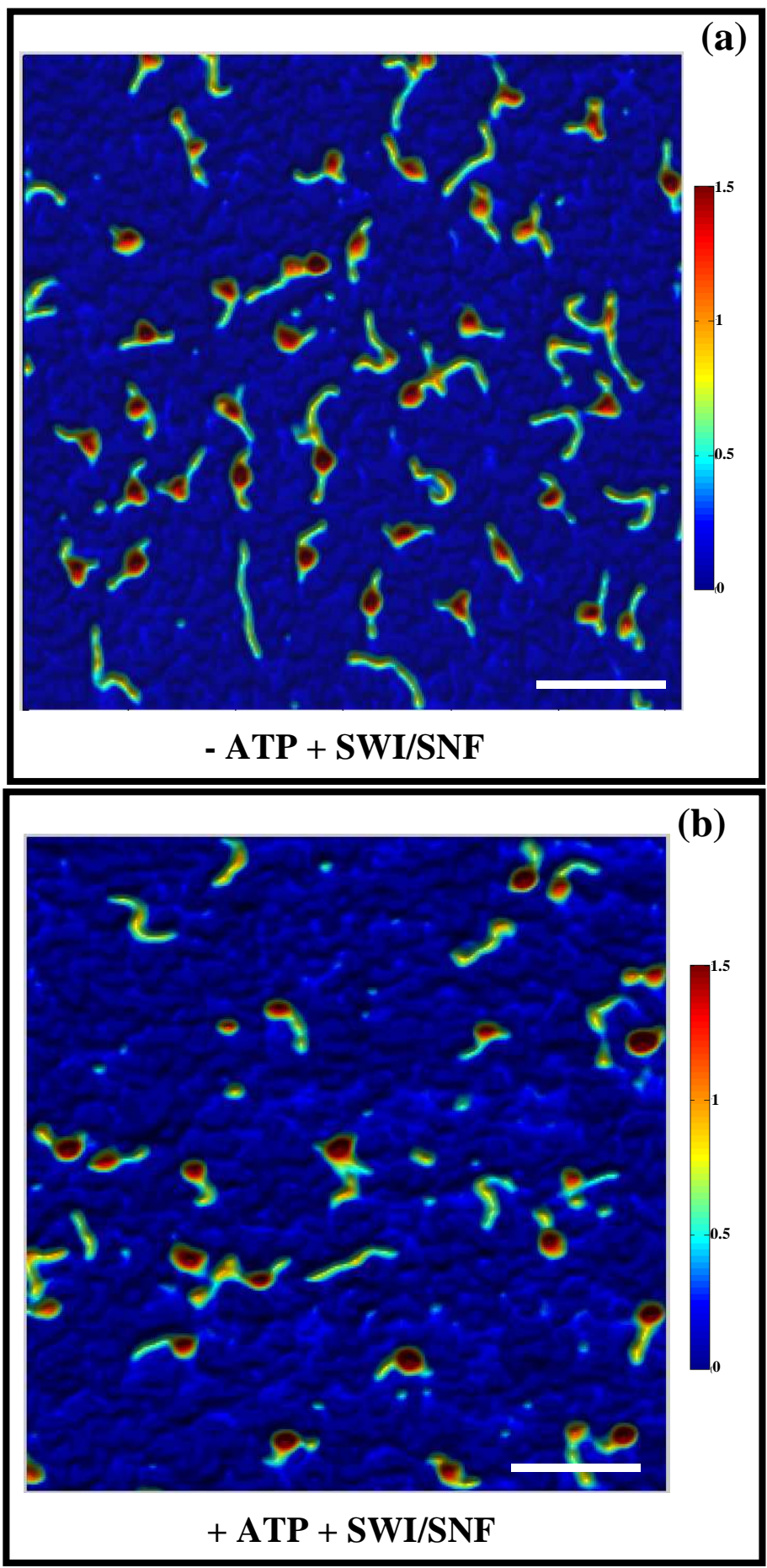
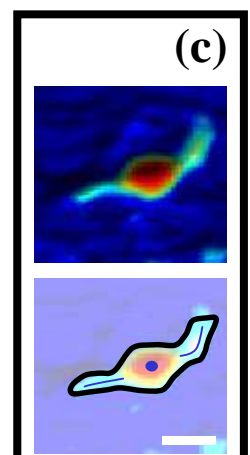
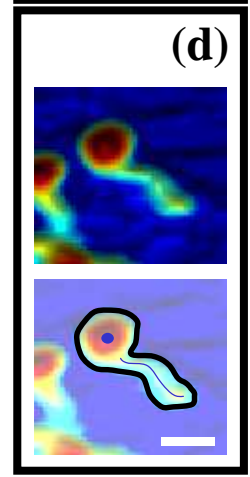

**Figure 3**

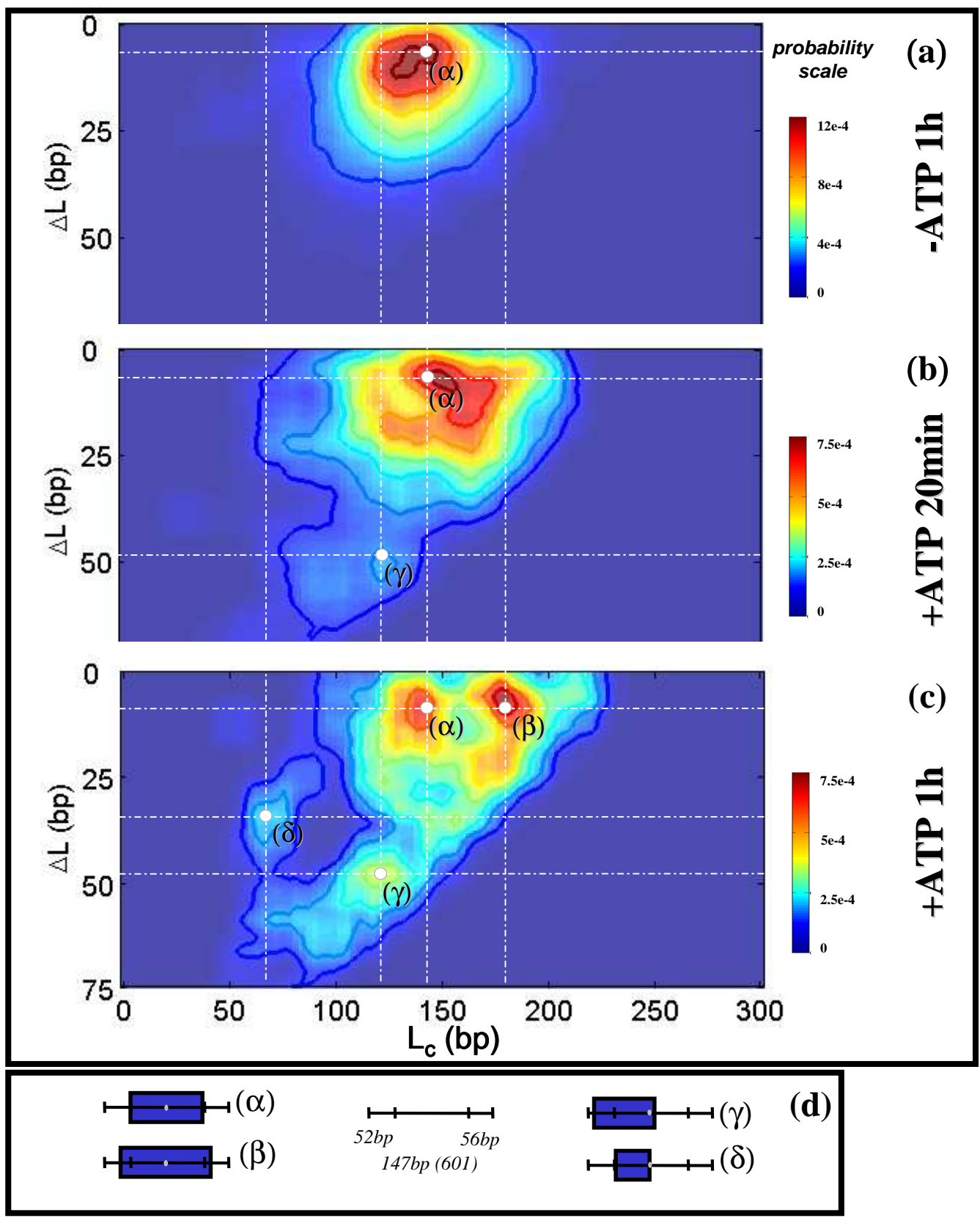

**Figure 4**

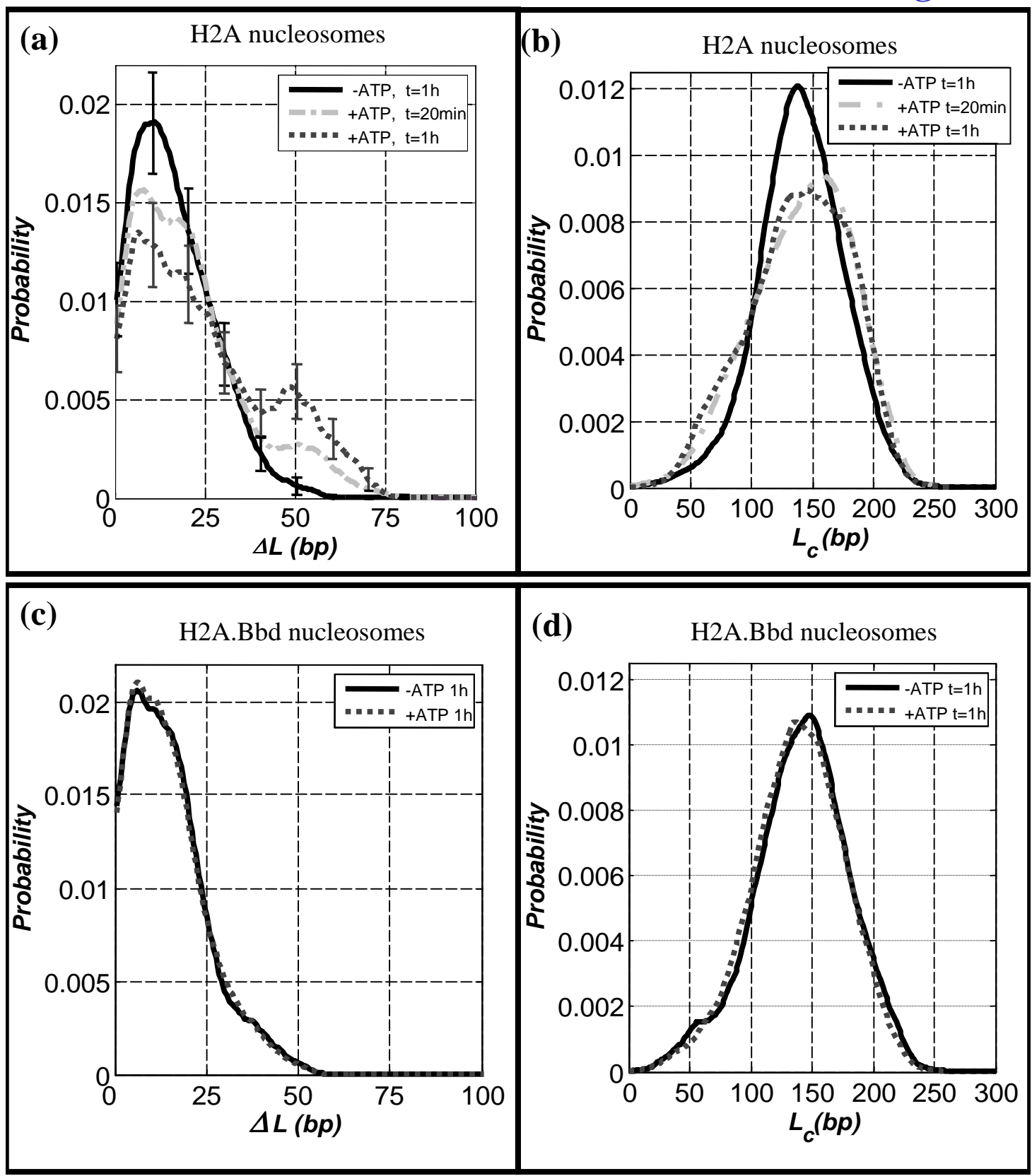

Figure 5